\documentclass[preprint,showpacs,preprintnumbers,amsmath,amssymb,eqsecnum]{revtex4}
\usepackage{graphicx}
\usepackage{dcolumn}
\usepackage{bm}

\newcommand{\bn}{\mbox{\boldmath $n$}}
\newcommand{\ba}{\mbox{\boldmath $a$}}

\newcommand{\be}{\mbox{\boldmath $e$}}
\newcommand{\bff}{\mbox{\boldmath $f$}}

\newcommand{\bA}{\mbox{\boldmath $A$}}
\newcommand{\bB}{\mbox{\boldmath $B$}}

\newcommand{\bJ}{\mbox{\boldmath $J$}}
\newcommand{\bK}{\mbox{\boldmath $K$}}

\newcommand{\bx}{\mbox{\boldmath $x$}}
\newcommand{\by}{\mbox{\boldmath $y$}}
\newcommand{\bz}{\mbox{\boldmath $z$}}
\newcommand{\bmm}{\mbox{\boldmath $m$}}
\begin{document}



\title{Decomposition of meron configuration of $SU(2)$ gauge field\\}

\author{Minoru Hirayama}
 \email{hirayama@sci.toyama-u.ac.jp}
\author{Jun Yamashita}
\email{fxjun@jodo.sci.toyama-u.ac.jp}
\affiliation{Department of Physics, Toyama University, Gofuku 3190, Toyama 930-8555, {\bf Japan}\\}


\date{November, 8, 2003}

\begin{abstract}
For the meron configuration of the $SU(2)$ gauge field in the four dimensional Minkowskii spacetime, the decomposition into an isovector field $\bn$, isoscalar fields $\rho$ and $\sigma$, and a $U(1)$ gauge field $C_{\mu}$ is attained by solving the consistency condition for $\bn$. The resulting $\bn$ turns out to possess two singular points, behave like a monopole-antimonopole pair and reduce to the conventional hedgehog in a special case. The $C_{\mu}$ field also possesses singular points, while $\rho$ and $\sigma$ are regular everywhere.
\begin{center}
\end{center}
\end{abstract}

\pacs{11.10.Lm,02.30.Ik,12.39.Dc}

\maketitle


\section{\label{sec:Introduction}Introduction\protect}

To describe the low energy behavior of the $SU(2)$ gauge field in the four dimensional Minkowskii spacetime $M^4$, several authors \cite{Cho}, \cite{HayashiMorita}, \cite{FaddeevNiemi} advocated to decompose the gauge-fixed $SU(2)$ gauge field $\bA_{\mu}(x)$ into some scalar fields and a $U(1)$ gauge field. In the notation of Faddeev and Niemi \cite{FaddeevNiemi}, $\bA_{\mu}(x)$ is decomposed as
\begin{align}
\bA_{\mu}=C_{\mu}\bn+\rho\partial_{\mu}\bn+\left(1+\sigma\right)\partial_{\mu}\bn\times\bn,
\label{eqn:FNdecomposition}
\end{align}
where $\bn(x)$ is an isovector scalar field satisfying
\begin{align}
\bn^2=\sum_{a=1}^3\left(n^a\right)^2=1,
\end{align}
$\rho(x)$ and $\sigma(x)$ are isoscalar scalar fields and $C_{\mu}(x)$ is a gauge-fixed $U(1)$ gauge field. When a configuration $\bA_{\mu}(x)$ and the field $\bn(x)$ are fixed, the fields $C_{\mu}(x)$, $\rho(x)$ and $\sigma(x)$ are given by
\begin{align}
&C_{\mu}=\bn\cdot\bA_{\mu}, \label{eqn:Cmu}\\
&\rho=\frac{\partial_{\mu}\bn\cdot\bA_{\mu}}{(\partial_{\mu}\bn)^2}, \label{eqn:rho}
\end{align} 
and
\begin{align}
1+\sigma=\frac{(\partial_{\mu}\bn\times\bn)\cdot\bA_{\mu}}{(\partial_{\mu}\bn\times\bn)^2},
\label{eqn:sigma}
\end{align}
where the suffix $\mu$ on the right hand sides of Eqs. (\ref{eqn:rho}) and (\ref{eqn:sigma}) should be fixed as 0 or 1 or 2 or 3 instead of being summed over $0\sim 3$. In other words, the field $\bn(x)$ must be chosen so as to ensure that $\partial_{\mu}\bn\cdot\bA_{\mu}/(\partial_{\mu}\bn)^2$ and $(\partial_{\mu}\bn\times\bn)\cdot\bA_{\mu}/(\partial_{\mu}\bn\times\bn)^2$ do not depend on $\mu$. It was shown \cite{HUKY} that this condition for $\bn(x)$ can be expressed as the following equation:
\begin{align}
&\begin{pmatrix}\partial_{\mu}\theta \\ \sin\theta~\partial_{\mu}\varphi \end{pmatrix}=\begin{pmatrix}\beta & \gamma \\ -\gamma & \beta \end{pmatrix}\begin{pmatrix}G_{\mu} \\ H_{\mu} \end{pmatrix} \label{eqn:Hirayamacondition}\\
&G_{\mu}=\frac{\partial\bn}{\partial\theta}\cdot\bA_{\mu},\quad H_{\mu}=\frac{1}{\sin\theta}\frac{\partial\bn}{\partial\varphi}\cdot\bA_{\mu},
\end{align}
where $\theta(x)$ and $\varphi(x)$ are defined by
\begin{align}
\bn(x)=\left(\sin\theta\cos\varphi,~\sin\theta\sin\varphi,~\cos\theta\right)
\label{eqn:parametrizen}
\end{align}
and $\beta(x)$ and $\gamma(x)$ are arbitrary functions. 

It is known that the static hedgehog configuration $\bn(x)=\bn(x_0, \bx)=\bx/r,~r=|\bx|$, can be used for the meron and the instanton configurations of $\bA_{\mu}(x)$ \cite{Konishi}, \cite{Tsutsui}. For example, Witten's $Ansatz$ \cite{Witten}
\begin{align}
&A_0^a=A_0\frac{x_a}{r}, \\
&A_k^a=\frac{\varphi_2+1}{r^2}\varepsilon_{kaj}x^j+\frac{\varphi_1}{r^3}\left(\delta_{ka}r^2-x^kx^a\right)+A_1\frac{x^kx^a}{r^2}
\end{align}
for instantons is compatible with the decomposition (\ref{eqn:FNdecomposition}) with $\bn(x)=\bx/r$ \cite{Tsutsui}.

The purpose of this paper is to explore possible configurations of $\bn(x)$ other than the hedgehog. We seek the solutions of Eq. (\ref{eqn:Hirayamacondition}) with $\bA_{\mu}$ being set as the meron which is regular everywhere in $M^4$. With a simplifying assumption for the pair of arbitrary functions $\beta(x)$ and $\gamma(x)$, we obtain the solution
\begin{align}
\bn(x)=\frac{2\bx\times\ba-\left(1+x^{\mu}x_{\mu}\right)\ba+2a_4\bx}{\left|2\bx\times\ba-\left(1+x^{\mu}x_{\mu}\right)\ba+2a_4\bx\right|},
\label{eqn:solutionn}
\end{align}
where $a=(\ba,~a_4)=(a_1,~a_2,~a_3,~a_4)$ is an arbitrary real 4-vector. The above $\bn(x)$ is a generalization of the hedgehog since it is equal to sgn$(a_4)\bx/r$ for vanishing $\ba$. For generic $a$ and fixed $x_0$, our solution (\ref{eqn:solutionn}) possesses a pair of singular points, say, $\bz_1(x_0)$ and $\bz_2(x_0)$. If we define the monopole charge associated with the singular point $\bz$ of $\bn(x)$ by
\begin{align}
M_{\bz}=\frac{1}{8\pi}\int_{S^2(\bz)}\bn\cdot\left(\partial_i\bn\times\partial_j\bn\right)dx^i\wedge dx^j
\end{align}
with $S^2(\bz)$ being a sphere surrounding the singular point $\bz$, we see $M_{\bz_1(x_0)}=-\mbox{sgn}(a_4)$ and $M_{\bz_2(x_0)}=\mbox{sgn}(a_4)$. Hence our solution describes a moving pair of a monopole and an antimonopole. For the special case of $\ba=0$, the singularity $\bz_1(x_0)$ goes away to the spatial infinity, leaving a single singularity with $M=\mbox{sgn}(a_4)$ at the origin. The singularity structure of the fields $C_{\mu}, \rho$ and $\sigma$ for generic $(\ba,~a_4)$ is also investigated. It is observed that $C_{\mu}$ is singular while $\rho$ and $1+\sigma$ vanish at $\bz_1(x_0)$ and $\bz_2(x_0)$.

This paper is organized as follows. In Sec. II, we describe the procedure to solve Eq. (\ref{eqn:Hirayamacondition}) and we investigate the singularity and the preimage of $\bn(x)$. In Sec. III, the behavior of $\rho, 1+\sigma$ and $C_{\mu}$ is investigated. The final section, Sec. IV is devoted to a summary and discussions.


\section{\label{sec:Field bn(x) for meron}Field $\bn(x)$ for meron\protect}

\subsection{Derivation of Eq. (\ref{eqn:Hirayamacondition})}

For self-containedness, we first discuss briefly how Eq. (\ref{eqn:Hirayamacondition}) is derived from the conditions (\ref{eqn:rho}) and (\ref{eqn:sigma}) \cite{HUKY}. Noting that the vectors $\bn$,
\begin{align}
\be\equiv\frac{\partial\bn}{\partial\theta}=\left(\cos\theta\cos\varphi,~\cos\theta\sin\varphi,~-\sin\theta\right)
\end{align}
and
\begin{align}
\bff\equiv\frac{1}{\sin\theta}\frac{\partial\bn}{\partial\varphi}=\left(-\sin\varphi,~\cos\varphi,~0\right)
\end{align}
constitute a right-handed orthonormal system and that
\begin{align}
\left(\partial_{\mu}\bn\right)^2&=\left(\partial_{\mu}\bn\times\bn\right)^2 \\
&=\left|\begin{matrix}\partial_{\mu}\theta & -\sin\theta~\partial_{\mu}\varphi \\ \sin\theta~\partial_{\mu}\varphi & \partial_{\mu}\theta \end{matrix}\right|,
\end{align}
Eqs. (\ref{eqn:rho}) and (\ref{eqn:sigma}) are rewritten as
\begin{align}
M_{\mu}\begin{pmatrix}G_{\nu} \\ H_{\nu}\end{pmatrix}=M_{\nu}\begin{pmatrix}G_{\mu} \\ H_{\mu}\end{pmatrix}
\end{align}
with
\begin{align}
M_{\mu}=\begin{pmatrix}\partial_{\mu}\theta & -\sin\theta~\partial_{\mu}\varphi \\ \sin\theta~\partial_{\mu}\varphi & \partial_{\mu}\theta \end{pmatrix}.
\end{align}
With the help of the relations
\begin{align}
&\begin{pmatrix} a & -b \\ b & a \end{pmatrix}\begin{pmatrix} u \\ v \end{pmatrix}=\begin{pmatrix} u & -v \\ v & u \end{pmatrix}\begin{pmatrix} a \\ b \end{pmatrix}, \\
&\begin{pmatrix} a & -b \\ b & a \end{pmatrix}\begin{pmatrix} a^{\prime} & -b^{\prime} \\ b^{\prime} & a^{\prime} \end{pmatrix}=\begin{pmatrix} a^{\prime} & -b^{\prime} \\ b^{\prime} & a^{\prime} \end{pmatrix}\begin{pmatrix} a & -b \\ b & a \end{pmatrix},  \\
& a,~b,~a^{\prime},~b^{\prime},~u,~v\in\mathbb{C},
\end{align}
we obtain
\begin{align}
\begin{pmatrix}G_{\mu} & -H_{\mu} \\ H_{\mu} & G_{\mu} \end{pmatrix}^{-1}\begin{pmatrix}\partial_{\mu}\theta \\ \sin\theta\partial_{\mu}\varphi\end{pmatrix}=\begin{pmatrix}G_{\nu} & -H_{\nu} \\ H_{\nu} & G_{\nu} \end{pmatrix}^{-1}\begin{pmatrix}\partial_{\nu}\theta \\ \sin\theta\partial_{\nu}\varphi\end{pmatrix}.
\end{align}
Since the l. h. s. (r. h. s.) of the above equation should not depend on $\mu$ ($\nu$), we have
\begin{align}
\begin{pmatrix}\partial_{\mu}\theta \\ \sin\theta\partial_{\mu}\varphi\end{pmatrix}=\begin{pmatrix}G_{\mu} & -H_{\mu} \\ H_{\mu} & G_{\mu} \end{pmatrix}\begin{pmatrix}\beta \\ -\gamma \end{pmatrix},
\end{align}
which is equivalent to (\ref{eqn:Hirayamacondition}), where $\beta(x)$ and $\gamma(x)$ are arbitrary functions independent of $\mu$.


\subsection{Meron configuration}

To obtain the meron configuration \cite{dAFF} of the $SU(2)$ Yang-Mills field, L\"uscher \cite{Luscher} fully utilized the conformal invariance of the model. Since the conformal group of $M^4$ is isomorphic to $O(4, 2)$, we introduce the coordinates
\begin{align}
\xi&=\left(\xi^0,~\xi^1,~\xi^2,~\xi^3,~\xi^4,~\xi^5\right) \\
&=R\left(\sin\tau,~u^1,~u^2,~u^3,~u^4,~\cos\tau\right)
\end{align}
satisfying
\begin{align}
(\xi^0)^2-(\xi^1)^2-(\xi^2)^2-(\xi^3)^2-(\xi^4)^2+(\xi^5)^2=0.
\end{align}
They are related to $x^{\mu}$ by $x^{\mu}=\xi^{\mu}/\left(\xi^4+\xi^5\right)$, that is,
\begin{align}
x^0=\frac{\sin\tau}{\cos\tau+u^4},\quad x^k=\frac{u^k}{\cos\tau+u^4},
\label{eqn:map}
\end{align} 
and $u^{\alpha}$ ($\alpha=1,2,3,4)$ satisfy
\begin{align}
\sum_{\alpha=1}^4\left(u^{\alpha}\right)^2=1.
\end{align}

The conformal transformation from $x^{\mu}$ to $x^{\prime \mu}=\xi^{\prime\mu}/\left(\xi^{\prime 4}+\xi^{\prime 5}\right)$ is realized by $\xi^{\prime A}=\Lambda^A_B\xi^B,~\left(A=0,1,2,3,4,5\right)$ with $\Lambda\in O(4,2)$. We define the 1-forms $\omega^{\mu}$ ($\mu=0, 1, 2, 3$) by
\begin{align}
\omega^0=d\tau,\quad \omega^k=\eta_{\alpha\beta}^k u^{\alpha}du^{\beta},
\end{align}
where $\eta_{\alpha\beta}^k$ is the 't Hooft symbol \cite{tHooft} defined by $\eta_{\alpha\beta}^k=-\eta_{\beta\alpha}^k,~\eta_{j4}^k=\delta_{kj}$ and $\eta_{ij}^k=\varepsilon_{ijk}$. If we define the differential operators $\nabla_{\mu}$ by
\begin{align}
\nabla_0=\frac{\partial}{\partial\tau},\quad \nabla_k=\frac{1}{2}\eta_{\alpha\beta}^k\left(u^{\alpha}\frac{\partial}{\partial u^{\beta}}-u^{\beta}\frac{\partial}{\partial u^{\alpha}}\right),
\end{align}
we are led to the formula
\begin{align}
df=\omega^{\mu}\nabla_{\mu}f
\label{eqn:generalf}
\end{align}
for an arbitrary function $f$ of $x$. The meron configuration is given by
\begin{align}
&A_0^a=-4\Gamma^2x^0x^a, \\
&A_k^a=4\Gamma^2\left[\frac{1}{2}\left(1+x_{\mu}x^{\mu}\right)\delta_{ak}+\varepsilon_{akj}x^j+x^ax^k\right], \\
&\Gamma=\left[\left(1+x_{\mu}x^{\mu}\right)^2+4\bx^2\right]^{-\frac{1}{2}}. \label{eqn:Gamma}
\end{align}
If we define $B_{\mu}^a$ by
\begin{align}
A_{\mu}^adx^{\mu}=B_{\mu}^a\omega^{\mu},
\label{eqn:AandB}
\end{align}
$B_{\mu}^a$ takes a very simple form \cite{Luscher}:
\begin{align}
B_0^a=0,\quad B_k^a=-\delta_{ak}.
\label{eqn:solB}
\end{align}


\subsection{Rewriting Eq. (\ref{eqn:Hirayamacondition})}

In this subsection, we rewrite Eq. (\ref{eqn:Hirayamacondition}) with the help of the operator $\nabla_{\mu}$. From the definitions of $\be$ and $\bff$, we have
\begin{align}
\partial_{\mu}n^k=e^k\partial_{\mu}\theta+\sin\theta f^k\partial_{\mu}\varphi.
\end{align}
Making use of Eq. (\ref{eqn:Hirayamacondition}), we obtain
\begin{align}\
\partial_{\mu}n^k=\left(\beta e^k-\gamma f^k\right)\left(\be\cdot\bA_{\mu}\right)+\left(\gamma e^k+\beta f^k\right)\left(\bff\cdot\bA_{\mu}\right).
\end{align}
From the relations (\ref{eqn:generalf}) and (\ref{eqn:AandB}), we find
\begin{align}
dn^k&=\omega^{\mu}\nabla_{\mu}n^k \\
&=\left(\beta e^k-\gamma f^k\right)\left(\be\cdot\bA_{\mu}dx^{\mu}\right)+\left(\gamma e^k+\beta f^k\right)\left(\bff\cdot\bA_{\mu}dx^{\mu}\right) \\
&=\left(\beta e^k-\gamma f^k\right)\left(\be\cdot\bB_{\mu}\omega^{\mu}\right)+\left(\gamma e^k+\beta f^k\right)\left(\bff\cdot\bB_{\mu}\omega^{\mu}\right)
\end{align}
and hence
\begin{align}
\nabla_{\mu}n^k=\left(\beta e^k-\gamma f^k\right)\left(\be\cdot\bB_{\mu}\right)+\left(\gamma e^k+\beta f^k\right)\left(\bff\cdot\bB_{\mu}\right).
\end{align}
With the help of Eq. (\ref{eqn:solB}), $\be\times\bff=\bn$ and $e^je^k+f^jf^k+n^jn^k=\delta^{jk}$, we find that Eq. (\ref{eqn:Hirayamacondition}) is equivalent to
\begin{align}
&\nabla_0n^k=0, \\
&\nabla_jn^k=\beta\left(n^jn^k-\delta^{jk}\right)+\gamma\varepsilon_{jkl}n^l.
\label{eqn:ModifiedHirayama}
\end{align}


\subsection{A solution of Eq. (\ref{eqn:ModifiedHirayama})}

To find a solution of Eq. (\ref{eqn:ModifiedHirayama}), we assume that $n^k(x)$ is of the following form:
\begin{align}
n^k=f(v)\nabla_k v,\quad v\equiv a\cdot u=\sum_{\alpha=1}^4 a_{\alpha}u^{\alpha},
\end{align}
where $a_{\alpha}~\left(\alpha=1,2,3,4\right)$ are real constants. Then we have
\begin{align}
\nabla_jn^k&=f^{\prime}(v)\left(\nabla_j v\right)\left(\nabla_k v\right)+f(v)\nabla_j \nabla_k v \label{eqn:nablan} \nonumber \\
&=\frac{f^{\prime}(v)}{\left[f(v)\right]^2}n^jn^k-\delta_{jk}vf(v)-\varepsilon_{jkl}n^l.
\end{align}
If $f(v)$ satisfies
\begin{align}
\frac{f^{\prime}(v)}{\left[f(v)\right]^2}=vf(v),
\label{eqn:solf}
\end{align}
the r. h. s. of Eq. (\ref{eqn:nablan}) is equal to $\beta(n^jn^k-\delta_{jk})+\gamma\varepsilon_{jkl}n^l$ with
\begin{align}
\beta=vf(v),\quad \gamma=-1.
\end{align}
The solution of Eq. (\ref{eqn:solf}) is given by
\begin{align}
f(v)=\frac{1}{\sqrt{b-v^2}},\quad b=\mbox{const.},
\end{align}
and hence $n^k$ is given by
\begin{align}
n^k=\frac{1}{\sqrt{b-v^2}}\eta_{\alpha\beta}^k u^{\alpha}a_{\beta}.
\end{align}
The constant $b$ is fixed by the condition $\bn^2=1$ as
\begin{align}
b=a^2\equiv\sum_{\alpha=1}^4\left(a_{\alpha}\right)^2.
\end{align}
Hence we obtain
\begin{align}
n^k=\frac{\eta_{\alpha\beta}^k u^{\alpha}a_{\beta}}{\sqrt{a^2-(a\cdot u)^2}}.
\end{align}
Since the inverse of the relation (\ref{eqn:map}) is given by
\begin{align}
u^k=2x^k\Gamma,\quad u^4=\left(1+x_{\mu}x^{\mu}\right)\Gamma
\end{align}
with $\Gamma$ defined by Eq. (\ref{eqn:Gamma}), we are led to
\begin{align}
\eta_{\alpha\beta}^k u^{\alpha}a_{\beta}=\Gamma\left[2\bx\times\ba-(1+x_{\mu}x^{\mu})\ba+2a_4\bx\right]^k
\end{align}
with $\bx=(x^1,~x^2~x^3)$ and $\ba=(a_1,~a_2,~a_3)$. Then we find that $\bn(x)$ is given by Eq. (\ref{eqn:solutionn}).


\subsection{Singularities of $\bn(x)$}

We have obtained $\bn(x)$ in the following form:
\begin{align}
&\bn=\frac{\bmm}{\left|\bmm\right|}, \label{eqn:nbym}\\
&\bmm=2\bx\times\ba-\left[1+(x_0)^2-\bx^2\right]\ba+2a_4\bx.
\label{eqn:bmm}
\end{align}
If we assume $\ba\neq 0$ and define $F_+(x_0), F_-(x_0), \kappa, \by_+$ and $\by_-$ by
\begin{align}
&F_+(x_0)=\sqrt{1+(x_0)^2+\kappa^2}+\kappa, \\
&F_-(x_0)=\sqrt{1+(x_0)^2+\kappa^2}-\kappa, \\
&\kappa=\frac{|a_4|}{|\ba|}, \\
&\by_+=\bx+\mbox{sgn}(a_4)F_+(x_0)\hat{\ba}
\end{align}
and
\begin{align}
\hspace{-8mm}\by_-=\bx-\mbox{sgn}(a_4)F_-(x_0)\hat{\ba},
\end{align}
$\bmm$ can be written as
\begin{align}
\bmm&=2\by_+\times\ba+\left(\left[\bx-\mbox{sgn}(a_4)F_+(x_0)\hat{\ba}\right]\cdot\by_+\right)\ba+2a_4\by_+ \\
&=2\by_-\times\ba+\left(\left[\bx+\mbox{sgn}(a_4)F_-(x_0)\hat{\ba}\right]\cdot\by_-\right)\ba+2a_4\by_-,
\end{align}
implying that $\bn(x)$ is singular when $\by_+$ or $\by_-$ vanishes. Thus we see that, for fixed $x_0$, our configuration (\ref{eqn:nbym}) is singular at
\begin{align}
\bz_1(x_0)=-\mbox{sgn}(a_4)F_+(x_0)\hat{\ba}
\end{align}
and
\begin{align}
\bz_2(x_0)=\mbox{sgn}(a_4)F_-(x_0)\hat{\ba}.
\end{align}
In the neighborhood of $\bz_1(x_0)$, we have
\begin{align}
\bmm\sim 2\left[\by_+\times\ba+\left(\bz_1(x_0)\cdot\by_+\right)\ba+a_4\by_+\right]
\end{align}
or
\begin{align}
&\begin{pmatrix}m^1 \\ m^2 \\ m^3\end{pmatrix}\sim 2B \begin{pmatrix} y_+^1 \\ y_+^2 \\ y_+^3\end{pmatrix}, \\
&B=-\mbox{sgn}(a_4)F_+(x_0)\begin{pmatrix}\hat{a}_1a_1 &\hat{a}_1a_2 &\hat{a}_1a_3 \\ \hat{a}_2a_1 &\hat{a}_2a_2 &\hat{a}_2a_3 \\ \hat{a}_3a_1 &\hat{a}_3a_2 &\hat{a}_3a_3\end{pmatrix}+\begin{pmatrix} a_4 & a_3 & -a_2 \\ -a_3 & a_4 & a_1 \\ a_2 & -a_1 & a_4 \end{pmatrix}.
\end{align}
Then the configuration $\bn(x_0,~\bx)$ is a twisted hedgehog and hence the monopole charge associated with the singularity $\bz_1(x_0)$ is given by
\begin{align}
M_{\bz_1(x_0)}=\mbox{sgn}\left(\mbox{det}B\right).
\end{align}
Since $\mbox{det}B$ is calculated to be $-\mbox{sgn}(a_4)\sqrt{1+(x_0)^2+\kappa^2}|\ba|a^2,$ we have
\begin{align}
M_{\bz_1(x_0)}=-\mbox{sgn}(a_4).
\end{align}
Similarly the singularity $\bz_2(x_0)$ yields the monopole charge
\begin{align}
M_{\bz_2(x_0)}=\mbox{sgn}(a_4).
\end{align}
Since both $F_-(x_0)$ and $F_+(x_0)$ decrease with $x_0$ for $x_0<0$ and increase for $x_0>0$, the singularities $\bz_1(x_0)$ and $\bz_2(x_0)$ look like a moving pair of a monopole and an antimonopole.


\subsection{\label{sec:The limit}The $\ba\rightarrow 0$ limit\protect}

We here consider the monopole charge $M(R)$ inside the sphere which is centered at the origin and of radius $R$:
\begin{align}
M(R)=\frac{1}{8\pi}\int_{|\bx|=R}\bn\cdot\left(\partial_i\bn\times\partial_j\bn\right)dx^i\wedge dx^j.
\end{align}
We easily see that $M(R)$ is given by
\begin{align}
M(R)=\left\{\begin{aligned}
&0&; &R>F_+(x_0), \\
&\mbox{sgn}(a_4)&; & F_+(x_0)>R>F_-(x_0), \\
&0&; &F_-(x_0)>R.
\end{aligned}\right.
\end{align}

By rewriting the inequality $F_+(x_0)>R>F_-(x_0)$ as
\begin{align}
\frac{\sqrt{|\ba|^2\left(1+x_0^2\right)+(a_4)^2}+\left|a_4\right|}{|\ba|}>R>\frac{\left(1+x_0^2\right)|\ba|}{\sqrt{|\ba|^2\left(1+x_0^2\right)+(a_4)^2}+|a_4|},
\end{align}
we see that, in the $|\ba|\rightarrow 0$ limit, $M(R)$ is equal to sgn($a_4$) for $\infty>R>0$. Thus we have understood how the conventional hedgehog appears from our solution:
the point $\bz_1(x_0)$ goes away to the spatial infinity and the point $\bz_2(x_0)$ approaches to the origin in the $|\ba|\rightarrow 0$ limit.


\subsection{$\bn(x_0, \bx)$ for $\left|\bx\right|=\infty$ and $\bx\propto\ba$}

For fixed $x_0$ and nonvanishing $\ba$, the configuration (\ref{eqn:parametrizen}) satisfies the boundary condition
\begin{align}
\left.\bn(x_0,~\bx)\right|_{|\bx|=\infty}=\hat{\ba}.
\label{eqn:bCondition}
\end{align}
Since $\bmm$ for $\bx=\xi\hat{\ba}$ is given by
\begin{align}
\bmm(x_0,~\xi\hat{\ba})&=2\xi\hat{\ba}\times\ba-\left[1+(x_0)^2-\xi^2\right]\ba+2a_4\xi\hat{\ba} \\
&=\ba\left[\xi+\mbox{sgn}(a_4)F_+(x_0)\right]\left[\xi-\mbox{sgn}(a_4)F_-(x_0)\right],
\end{align}
we have
\begin{align}
\bn(x_0,~\xi\hat{\ba})=\left\{\begin{aligned}&\hat{\ba}: &&\xi>M, \\
-&\hat{\ba}: &&N<\xi <M, \\&\hat{\ba}: &&\xi<N, \end{aligned}\right.
\end{align}
where $M$ and $N$ are defined by
\begin{align}
&M=\mbox{Max}\left[-\mbox{sgn}(a_4)F_+(x_0),\quad \mbox{sgn}(a_4)F_-(x_0)\right], \\
&N=\mbox{Min}\left[-\mbox{sgn}(a_4)F_+(x_0),\quad \mbox{sgn}(a_4)F_-(x_0)\right].
\end{align}
From the above, the configuration $\bn(x_0,~\bx)$ for $|\bx|=\infty$ and $\bx=\xi\hat{\ba}$ with $a_4>0$ can be illustrated as in Fig. 1
\begin{center}
\begin{figure}[hbtp]
\scalebox{1}[1]{\includegraphics{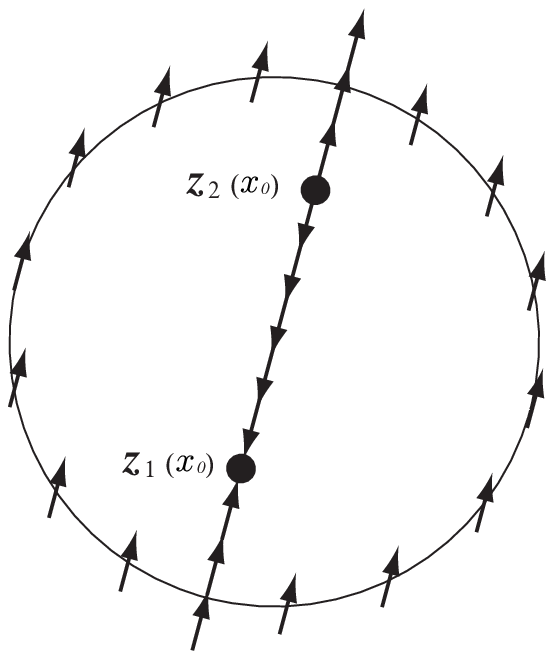}}
\label{figure:dipole}
\caption{Configuration $\bn$ for $|\bx|=\infty$ and $\bx\propto\ba$}
\end{figure}
\end{center}


\subsection{\label{sec:Preimage of n}Preimage of \bn \protect}

We see that $\bn(x_0,~\bx)$ with $x_0$ fixed is defined on the space $S^3\backslash \left\{\bz_1(x_0),~\bz_2(x_0)\right\}\equiv T$, where $S^3$ is the compactified $\mathbb{R}^3$. Since $T$ is not equal to $S^3$, we cannot regard $\bn$ as a mapping from $S^3$ to $S^2$ and we cannot define the Hopf charge for our $\bn(x_0,~\bx)$. In the case that $\bn$ is regular for $|\bx|<\infty$ and satisfies the boundary condition (\ref{eqn:bCondition}), the Hopf charge of $\bn$ is defined as the linking number of two preimages (loops) of $\bn$ in $S^3$. To understand the present situation more explicitly, we consider the preimage of $\bn(x_0,~\bx)$. It can be seen to be equal to
\begin{align}
\begin{aligned}
{\cal P}(\bn)=&\left\{\bx^{\prime}\left|\bmm(x_0,~\bx^{\prime})=\lambda\bmm(x_0,~\bx)~(\lambda\geq 0),\right.\right. \\
&\left.\lim_{\bx^{\prime}\rightarrow\bz_1(x_0)}\bmm(x_0,~\bx^{\prime})=\lim_{\bx^{\prime}\rightarrow\bz_2(x_0)}\bmm(x_0,~\bx^{\prime})=\bn(x_0,~\bx)\right\}.
\end{aligned}
\end{align}
The solution of the equation $\bmm(x_0,~\bx^{\prime})=\lambda\bmm(x_0,~\bx)$ can be obtained by setting $\bx^{\prime}=f\ba+g\bx+h\ba\times\bx$ with $f, g$ and $h$ being functions of $|\bx|$ and $x_0$. It is easily found that $f, g$ and $h$ are given by
\begin{align}
&\left(f\ba^2+\lambda\ba\cdot\bx+a_4\right)^2=\ba^2(1-\lambda)\left(1+x_0^2+\lambda\bx^2\right)+\left(\lambda\ba\cdot\bx+a_4\right)^2, \nonumber\\
&\left.g\right.=\lambda,\quad h=0.
\end{align}
Thus we find that ${\cal P}(\bn)$ consists of $\bx^{\prime}_+(\lambda)$ and $\bx^{\prime}_-(\lambda)$ with $\lambda_1\geq\lambda\geq 0$, where $\bx^{\prime}_+(\lambda)$, $\bx^{\prime}_-(\lambda)$ and $\lambda_1$ are defined by
\begin{align}
&\bx^{\prime}_+(\lambda)=\lambda\bx+f_+(\lambda)\ba, \\
&\bx^{\prime}_-(\lambda)=\lambda\bx+f_-(\lambda)\ba, \\
&f_{\pm}(\lambda)=-\frac{\lambda\ba\cdot\bx+a_4}{\ba^2}\left[1\pm\sqrt{1+\frac{(1-\lambda)\ba^2(1+x_0^2+\lambda\bx^2)}{(\lambda\ba\cdot\bx+a_4)^2}}\right], \\
&\left(\lambda\ba\cdot\bx+a_4\right)^2+(1-\lambda)\ba^2(1+x_0^2+\lambda\bx^2)=-(\ba\times\bx)^2(\lambda-\lambda_1)(\lambda-\lambda_2)
\end{align}
with $\lambda_1>\lambda_2$. More explicitly, $\lambda_1$ and $\lambda_2$ are given by
\begin{align}
\lambda_{1, 2}=&\frac{a_4(\ba\cdot\bx)+\frac{\ba^2}{2}\left[\bx^2-1-(x_0)^2\right]}{\left(\ba\times\bx\right)^2}\pm \\
&\sqrt{\cfrac{(a_4)^2+\ba^2\left[1+(x_0)^2\right]}{\left(\ba\times\bx\right)^2}+\frac{\left\{a_4(\ba\cdot\bx)+\frac{\ba^2}{2}\left[\bx^2-1-(x_0)^2\right]\right\}^2}{\left[\left(\ba\times\bx\right)^2\right]^2}}.
\end{align}
The preimage ${\cal P}(\bn)$ is depicted in Fig. 2
\begin{figure}[hbtp]
\scalebox{1}[1]{\includegraphics{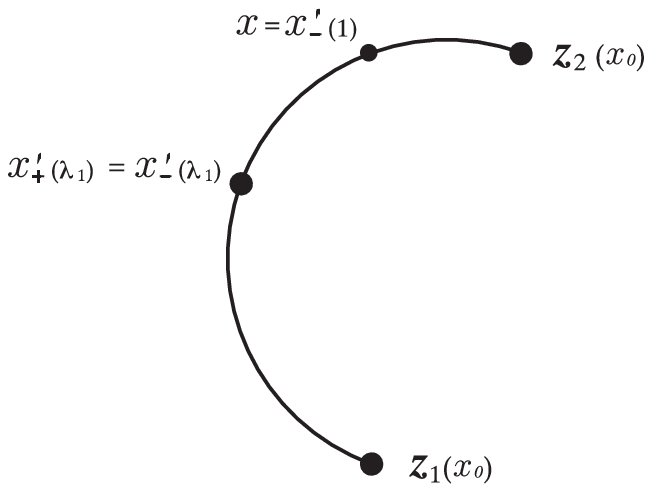}}
\label{figure:preimage1}
\caption{Preimage ${\cal P}(\bn)$}
\end{figure}

From the above discussion we obtain the configuration $\bn(x_0,~\bx)$ as is depicted in Fig. 3.
\begin{figure}[hbtp]
\scalebox{1}[1]{\includegraphics{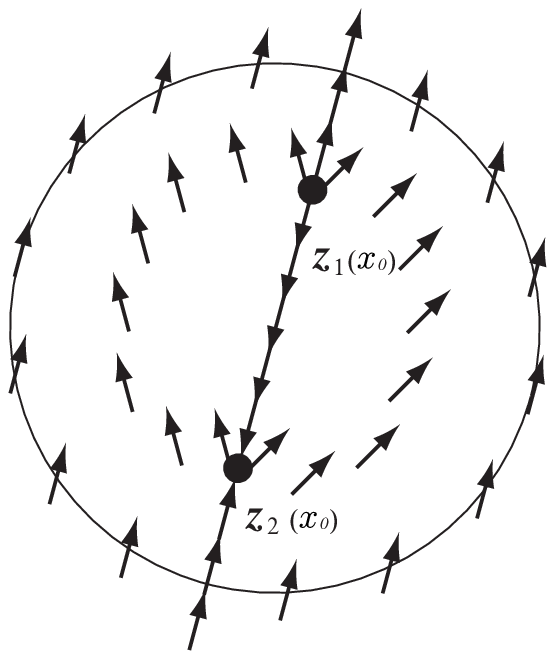}}
\caption{Configuration $\bn(x_0, \bx)$}
\label{figure:dipole2}
\end{figure}
\newpage

We note that the sum ${\cal P}(\bn)\cup{\cal P}(-\bn)$ consists of $\bx^{\prime}_+(\lambda)$ and $\bx^{\prime}_-(\lambda)$ with $\lambda_1\geq\lambda\geq\lambda_2$ and equals the loop in Fig. 4.

\begin{figure}[hbtp]
\scalebox{1}[1]{\includegraphics{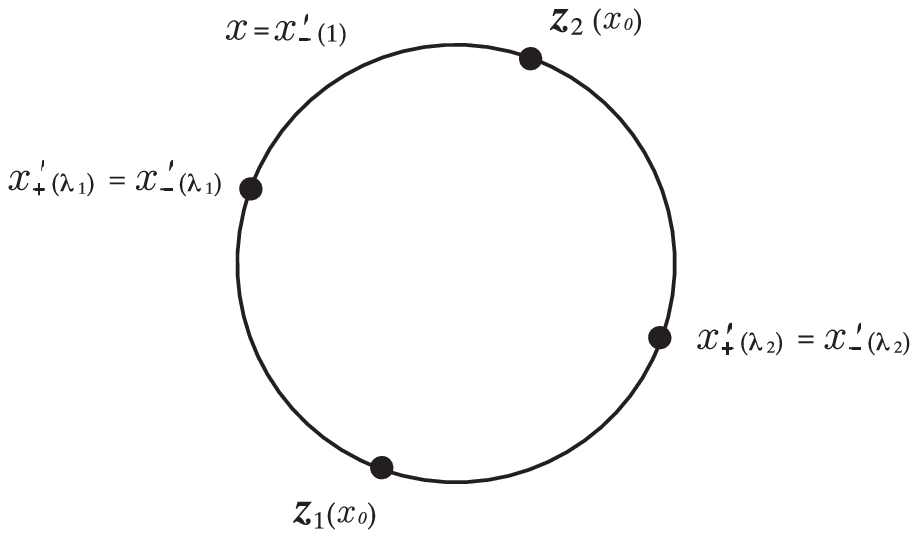}}
\label{figure:loop}
\caption{Premage ${\cal P}(\bn)\cup{\cal P}(-\bn)$}
\end{figure}

Then the two sets ${\cal P}(\bn)\cup{\cal P}(-\bn)$ and ${\cal P}(\bn^{\prime})\cup{\cal P}(-\bn^{\prime})$ become as in Fig. 5. They do not link but cross at $\bz_1(x_0)$ and $\bz_2(x_0)$ and hence their linking number cannot be defined.

\begin{figure}[hbtp]
\scalebox{1}[1]{\includegraphics{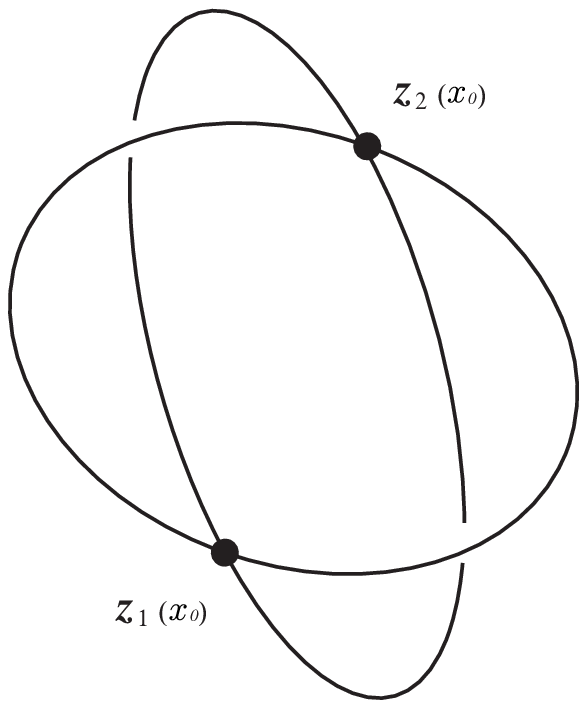}}
\label{figure:link}
\caption{Preiamges ${\cal P}(\bn)\cup{\cal P}(-\bn)$ and ${\cal P}(\bn^{\prime})\cup{\cal P}(-\bn^{\prime})$}
\end{figure}


\section{\label{sec:rho, sigma, and C_{mu}}$\rho, \sigma,$ and $C_{\mu}$\protect}

After straightforward but tedious calculations, we obtain
\begin{align}
&\rho=\frac{\partial_0\bn\cdot\bA_0}{\left(\partial_0\bn\right)^2}=\frac{\partial_k\bn\cdot\bA_k}{\left(\partial_k\bn\right)^2}=\frac{\sqrt{H}\xi}{H+\xi^2} \label{eqn:tablerho}\\
&1+\sigma=\frac{\left(\partial_0\bn\times\bn\right)\cdot\bA_0}{\left(\partial_0\bn\times\bn\right)^2}=\frac{\left(\partial_k\bn\times\bn\right)\cdot\bA_k}{\left(\partial_k\bn\times\bn\right)^2}=\frac{H}{H+\xi^2} \label{eqn:tablesigma}\\
&C_0=\bn\cdot\bA_0=\frac{4\Gamma^2}{\sqrt{H}}\eta x^0 \\
&C_k=\bn\cdot\bA_k=\frac{4\Gamma^2}{\sqrt{H}}\left[\left(\xi-\eta\right)x^k-\frac{a_k}{2\Gamma^2}\right], \\
&\begin{aligned}
H&=\bmm^2 \\
&=a^2\left[(1+x_{\mu}x^{\mu})^2+4\bx^2\right]-\left[a_4(1+x_{\mu}x^{\mu})+2\ba\cdot\bx\right]^2, \\
\xi&=a_4(1+x_{\mu}x^{\mu})+2\ba\cdot\bx, \\
\eta&=(1+x_{\mu}x^{\mu})(\ba\cdot\bx)-2a_4\bx^2,
\end{aligned}
\end{align}
where $\Gamma$ and $\bn$ are given by Eqs. (\ref{eqn:Gamma}) and (\ref{eqn:bmm}), respectively. We note that our $\bn$ indeed has the property mentioned below (\ref{eqn:Hirayamacondition}). We see that, at the points $\bz_1(x_0)$ and $\bz_2(x_0)$, the components $C_0(x)$ and $C_k(x)$ blow up while the fields $\rho(x)$ and $1+\sigma(x)$ vanish. The behavior of these fields at $\bx=0$, $\bz_i(x_0)~(i=1,2)$ and the spatial infinity is summarized in Table 1.
\[\begin{array}{|c|c|c|c|}\hline
&\left|\bx\right|=0 &\bx=\bz_1(x_0),\bz_2(x_0)&\left|\bx\right|=\infty \\
\hline
\bn &-\hat{\ba} &\mbox{indeterminate} &\hat{\ba} \\
C_0 &0 &\mbox{indeterminate} &0 \\
C_k &-\cfrac{2a_k}{|\ba|\left[1+(x_0)^2\right]} &\mbox{indeterminate} & 0 \\
\rho&\cfrac{a_4|\ba|}{a^2} &0 &-\cfrac{a_4|\ba|}{a^2} \\
1+\sigma &\cfrac{\ba^2}{a^2} &0 &\cfrac{\ba^2}{a^2} \\
\hline \end{array}\]
\begin{center}
Table 1
\end{center}


\section{\label{sec:Summary and discussions}Summary and discussion\protect}

We have found that the configurations $\bn(x_0, \bx)$ given by Eq. (\ref{eqn:solutionn}) can be the $\bn$ field of the decomposition (\ref{eqn:FNdecomposition}) of the meron configuration of the $SU(2)$ gauge field. For generic $a=(\ba, a_4)$, it possesses two singular points, behaves like a monopole-antimonopole pair and satisfies the boundary condition (\ref{eqn:bCondition}). The distance between the monopole and the antimonopole for fixed $x_0$ is given by $d=|\bz_1(x_0)-\bz_2(x_0)|=2\sqrt{(x_0)^2+(a^2/\ba^2)}$, where $a^2$ is equal to $(a_4)^2+\ba^2$. For vanishing $\ba$, the distance $d$ becomes infinity and the monopole or the antimonopole goes away to the spatial infinity. Although the case that the antimonopole disappears was discussed in II. E, it is clear that the opposite case in which the monopole is brought to the spatial infinity while the antimonopole is fixed at a finite point is possible. Thus we understand that, for a single configuration of $\bA_{\mu}$, various configurations of $\bn$ are allowed. The hedgehog is merely one example. Since the $\bn$ field corresponding to instanton solutions is usually identified with the hedgehog \cite{Tsutsui}, we expect that there are various configurations of $\bn$ also for an instanton. Other features of our $\bn(x_0, \bx)$ can be easily seen. For example, it approaches to $-\hat{\ba}$ and $\hat{\ba}$ in the $x_{\mu}x^{\mu}\gg 1$ and $x_{\mu}x^{\mu}\ll -1$ limits, respectively.

We have also investigated the behavior of $C_{\mu}(x), \rho(x)$ and $1+\sigma(x)$ and obtained the results given in Table 1. We here note two interesting properties of our solution. First, from the results (\ref{eqn:tablerho}) and (\ref{eqn:tablesigma}), we obtain the simple relation ${\rho}^2+(\sigma+1/2)^2=(1/2)^2$. Second, near the singularities $\bz_1(x_0)$ and  $\bz_2(x_0)$, the function $1+\sigma(x)$ vanishes as $|\bx-\bz_i(x_0)|^2$ while $\rho(x)$ as $|\bx-\bz_i(x_0)|$. Then in the neighborhood of $\bz_1(x_0)$ and $\bz_2(x_0)$, Eq. (\ref{eqn:FNdecomposition}) can be replaced by $\bA_{\mu}=C_{\mu}\bn+\rho\partial_{\mu}\bn$.

The appearance of the monopole-antimonopole pair in $\bn$ is reminiscent of the case of calorons which are periodic instanton solutions for the finite temperature $SU(N)$ gauge theory found by Kraan and Baal \cite{Kraan}. They showed that the caloron with a unit instanton number consists of $N$ basic BPS monopoles whose magnetic charges cancel exactly.  It was shown explicitly in Refs. \cite{Kraan, Bruckman} that the action densities of the $SU(2)$ and $SU(3)$ gauge fields in four dimensional euclidean space have two and three lumps in respective cases. Although it is beyond the scope of the present paper, it would be interesting to investigate if the $SU(N)$ version of the meron, if any, consists of $N$ monopoles.  On the other hand, the relation between the Hopf invariant and the instanton number was discussed by Taubes\cite{Taubes} and Jahn\cite{Jahn}. Jahn showed that the topological invariant, which he called the generalized Hopf invariant, can be defined for a  class of mappings $S^2 \times S^1 \rightarrow S^2$ and can be meaningful even if the Higgs field has singularities. In our case, $\bn(x)$ is not periodic in $x_0$ and cannot be regarded as a mapping from $S^2 \times S^1$ to $S^2$. It is interesting, however, to ask if a topological number can be defined for our $\bn(x)$.

It was suggested that, in analogy with the lower dimensional models such as 2+1 Georgi-Glashow model and the 1+1 dimensional Schwinger model, the regulated meron pair might play an important role in the discussion of the confinement of the color charge \cite{Steele}. Since the above discussion implies that the gauge field for the meron can be expressed simply as $C_{\mu}+\rho\partial_{\mu}\bn$ near the singularities, the color confinement mechanism might be discussed in a different manner.

At the end of the paper, we investigate if our $\bn$ solves the field equation of the model proposed in \cite{FaddeevNiemi,Faddeev}. If we define $F_{\mu\nu}(x)$ by
\begin{align}
F_{\mu\nu}=-F_{\nu\mu}=\frac{1}{2}\bn\cdot\left(\partial_{\mu}\bn\times\partial_{\nu}\bn\right)
\end{align}
and assume that the dynamics of $\bn(x)$ is governed by the Lagrangian density \cite{Faddeev}
\begin{align}
{\cal L}=c_2(\partial_{\mu}\bn)(\partial^{\mu}\bn)-2c_4F_{\mu\nu}F^{\mu\nu},
\end{align}
the field equation of $\bn(x)$ is given by
\begin{align}
\bK\equiv\partial_{\mu}\left(c_2\bn\times\partial^{\mu}\bn-2c_4F^{\mu\nu}\partial_{\nu}\bn\right)=0,
\end{align}
which is equivalent to the three equations
\begin{align}
\bx\cdot\bK=\ba\cdot\bK=\left(\bx\times\ba\right)\cdot\bK=0
\label{eqn:xKaK}
\end{align}
for generic $\bx$ and $\ba$.

After lengthy calculations, however, we find that the $\bn$ field that we obtained satisfies
\begin{align}
&\left.\bK\right|_{\bn(x)=(\ref{eqn:solutionn})}=F(x, a)\bJ, \\
&\bJ=\left(\bx\times\ba\right)\times\ba+a_4\left(\bx\times\ba\right),
\end{align}
where $F(x, a)$ is a complicated function which is nonvanishing for general $x$ and $a$. Thus our $\bn(x)$ satisfies only $\ba\cdot\bK=0$ and $(\bx\times\ba+a_4\bx)\cdot\bK=0$.


\begin{acknowledgments}
The authors are grateful to S. Kiryu for discussions at the early stage of this work. The authors also thank Shinji Hamamoto, Takeshi Kurimoto and Chang-Guang Shi for discussions. This work was supported in part by a Japanese Grant-in-Aid for Scientific Research from the Ministry of Education, Culture, Sports, Science and Technology (No.13135211).
\end{acknowledgments}

\end{document}